\documentclass[12pt,a4paper]{article}
\usepackage{a4wide}
\usepackage{epsfig}

\begin{document}

\begin{flushright}
MITP/13-009\\
January 2013
\end{flushright}

\begin{center}
{\Large\bf The decays of on-shell and off-shell polarized gauge\\[7pt]
bosons into massive quark pairs at NLO QCD}
\footnote{Invited talk given by J.G.~K\"orner at the 20th International 
Symposium on Spin Physics,
September 17-22, 2012, Dubna Russia. To be published in the Proceedings.}

\vspace{24pt}
S.~Groote$^a$, J.G.~K\"orner $^b$ and P.~Tuvike$^a$ 
 
\vspace{12pt}

{\small 
$^a${\it F\"u\"usika Instituut, Tartu \"Ulikool, Tartu, Estonia}\\ [.2cm]
$^b${\it PRISMA Cluster of Excellence, Institut f\"ur Physik,\\
  Johannes-Gutenberg-Universit\"at, Mainz, Germany}}
\end{center}

\centerline{\bf Abstract}
We discuss the polar angle decay distribution in the decay of on-shell and
off-shell polarized $(W,Z)$ gauge bosons into massive quark pairs. In
particular for the off-shell decays in
$H\to(W,Z)+(W^\ast,Z^\ast)(\to q_1\bar q_2)$ it is important to keep the masses
of the charm and bottom quarks at their finite values since the scale of the
problem is not set by $m_{W,Z}^2$ but by the offshellness  of the gauge boson
which varies in the range $(m_1+m_2)^2\le q^2\le(m_H-m_{W,Z})^2$.

\section{\label{sec:intro}Introduction}
The polarization of $W$ and $Z$ bosons produced in electroweak processes is in
general highly nontrivial. One therefore has a rich phenomenology of
polarization effects in $(W,Z)$ production and decay which will be explored in
present and future high energy experiments. The polarization of the $W$ and
$Z$ bosons can be probed by decay correlations involving the momenta of the
final state leptons or quarks in the decays of the polarized $(W,Z)$ bosons.
Bottom and charm quark mass effects are moderate for on-shell decays but are
important for off-shell decays in particular in the vicinity of the threshold
where the off-shell value $q^2$ becomes comparable to the quark masses. 
\section{Angular decay distribution}
In the $W$ rest frame the angular decay distribution of a polarized on-shell 
spin-1 $W$ boson into a pair of fermions is given by
\begin{eqnarray}\label{Wtheta}
W(\theta)&\propto&\frac32\quad
\sum_{m,m'=0,\pm}\rho_{mm}\,d^1_{mm'}(\theta)
  \,d^1_{mm'}(\theta)\, H_{m'm'}\nonumber\\
    &=&\frac38\cos^2\theta\,(\rho_{++}-2\rho_{00}+\rho_{--})
  (H_{++}-2H_{00}+H_{--})\nonumber\\&&
  +\frac34\cos\theta(\rho_{++}-\rho_{--})\,( H_{++}- H_{--})
  \nonumber\\&&
  +\frac38\Big((\rho_{++}+2\rho_{00}+\rho_{--})
  ( H_{++}+2 H_{00}+ H_{--})-4\rho_{00} H_{00}\Big).
\end{eqnarray}
The $\rho_{mm}$ are the process dependent (unnormalized) density matrix 
elements of the spin-1 gauge boson and the $H_{mm}$ are the (unnormalized) 
universal polarized decay functions needed to analyze the polarization of the
gauge boson. The polar angle $\theta$ is the angle between the $z$ direction
defined by the production process and a $z'$ direction specified by the decay
process. When the $z'$ axis is defined by the antiquark in the decay
$W^+\to c\bar b$ the (unnormalized) $O(\alpha_s)$ analyzing polarized decay
functions $H_{mm}$ are given by
\begin{eqnarray}\label{Hmm}
H_{++}&=&8N_cq^2\bigg[\,1+\frac{\alpha_s}{6\pi}
  \Big(1+(\pi^2+16)m_{2}/{\scriptstyle\sqrt{q^{2}}}\Big)+\ldots\bigg],
\nonumber\\
H_{00}\,\,\,&=&8N_cq^2\bigg[\,0+\frac{\alpha_s}{6\pi}
  \Big(4-2\pi^2m_{2}/{\scriptstyle\sqrt{q^{2}}}\Big)+\ldots\bigg],\nonumber\\
H_{--}&=&8N_cq^2\bigg[\,0+\frac{\alpha_s}{6\pi}
  \Big(1+(\pi^2-16) m_{2}/{\scriptstyle\sqrt{q^{2}}}\Big)+\ldots\bigg],
\end{eqnarray}
(cf.\ \cite{Groote:2012xr}) where we have expanded the result up to
$O(m_2/\sqrt{q^2})$. Surprisingly the mass corrections to the NLO terms are
linear in the antiquark mass and carry rather large coefficients.\footnote{For
$Z$ decays one has to replace the $H_{mm}$ by
$(v_f^2H^{VV}_{mm}+a_f^2H^{AA}_{mm})$ etc.\ in (\ref{Wtheta}), where $v_f$
and $a_f$ are the weak coupling coefficients of the SM.}
 
From (\ref{Wtheta}) one can define a normalized decay distribution
$\widehat W(\theta)$ by replacing the density matrix elements and the
polarized decay functions in (\ref{Wtheta}) by their normalized forms
$\hat\rho_{mm}=\rho_{mm}/\sum_{m}\rho_{mm}$ and
$\hat H_{mm}=H_{mm}/\sum_{m}H_{mm}$. One then has
$\int d\cos\theta\,\widehat W(\theta)=1$.
 
The angular decay distribution (\ref{Wtheta}) and its normalized form
$\widehat W(\theta)$ are second-order polynomials in $\cos\theta$ and
therefore have the functional form of a parabola. At threshold
$q^2=(m_1+m_2)^2$ one has $H_{++}=H_{00}=H_{--}$ and the angular decay
distribution becomes flat with $\widehat W(\theta)=1/2$.

The normalized decay distribution $\widehat W(\theta)$ can be characterized
by the convexity parameter (see e.g.\ \cite{Groote:2012xr,Groote:2013xt})
\begin{equation}\label{convex}
c_f=\frac{d^2\widehat W(\theta)}{d(\cos\theta)^2}
  =\frac34(\hat\rho_{++}-2\hat\rho_{00}+\hat\rho_{--})
  (\hat H_{++}-2\hat H_{00}+\hat H_{--}).
\end{equation}
When $c_{f}$ is negative (positive) the angular decay distribution is
described by a downward (upward) open parabola. The convexity parameter
vanishes for the two cases (i) unpolarized gauge boson where
$\hat\rho_{++}=\hat\rho_{00}=\hat\rho_{--}$ and/or (ii) zero analyzing power
where $\hat H_{++}=\hat H_{00}=\hat H_{--}$.

\section{The sequential decay $t\to b+W^+(\to c\bar b)$}
\begin{figure}\begin{center}
\epsfig{figure=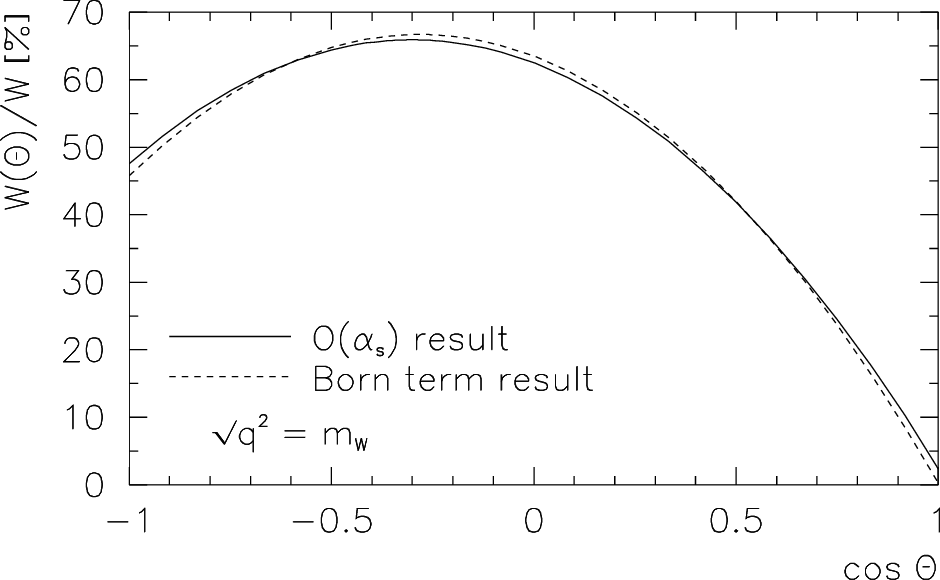,scale=0.8}
\caption{\label{nurks}Normalized angular decay distribution
  $\widehat W(\theta)=W(\theta)/W$ in the decay $W^+(\uparrow)\to c\bar b$ at
  LO (dashed line) and NLO (full line). NLO result contains both initial-state
  and final-state corrections.}
\end{center}\end{figure}

At LO the normalized density matrix elements of the $W^+$ in the decay
$t\to b+W^+$ are given by $\hat\rho_{++}=0$, $\hat \rho_{00}=(1+2x^2)^{-1}$
and $\hat\rho_{--}=2x^2/(1+2x^2)^{-1}$, where $x=m_W/m_t$~\cite{Kane:1991bg}.
The $z$ and $z'$ axes are defined by the momentum of the $W^+$ in the top rest
frame and the momentum of the $\bar b$-antiquark. NLO and NNLO corrections to
the density matrix elements $\rho_{mm}$ have been calculated in
\cite{Fischer:2001gp} and in \cite{Czarnecki:2010gb}, respectively. In
Fig.~\ref{nurks} we plot the normalized angular decay distribution for
$W^+(\uparrow)\to c\bar b$.
It is apparent that the distribution becomes flatter when radiative
corrections are applied. Numerically one has $c_f=-0.81$ and $c_f=-0.75$
without and with radiative corrections.
\section{ The sequential decays \\$H\to W^{-}+W^{\ast +}(\to q_1\bar{q}_2)$ 
and $H\to Z+Z^{\ast}(\to q\bar{q})$}
Quark mass effects are much more important for the off-shell decays since the
scale is not set by $m^2_{W,Z}$ but by the offshellness $q^2$ which varies in
the range $(m_1+m_2)^2\le q^2\le(m_H-m_{W,Z})^2$. There are also new scalar
contributions to the decay well familiar from neutron $\beta$ decay or e.g.\
from the decay $B\to(D,D^\ast)+\tau\nu_\tau$ where the scalar contributions
have been calculated to amount to $(67\,\%,37\,\%)$ of the total decay
rate~\cite{Korner:1989qb}. The scalar contributions can be isolated by
splitting the off-shell propagator of the $W^\ast,Z^\ast$ into a spin-1 and a
spin-0 piece by writing\footnote{In low energy applications the term
$q^2/M^2_{W,Z}$ in the scalar part is usually dropped.}
\begin{equation}\label{spin01}
\Big(-g^{\mu\mu'}+\frac{q^\mu q^{\mu'}}{m_{W,Z}^2}\Big)
  =\Big(-g^{\mu\mu'}+\frac{q^\mu q^{\mu'}}{q^2}
  -\frac{q^\mu q^{\mu'}}{q^2}(1-\frac{q^2}{m_{W,Z}^2})\Big).
\end{equation}

The angular decay distribution (\ref{Wtheta}) is now augmented by a scalar
contribution leading to
\begin{eqnarray}\label{offshell}
\lefteqn{W_{\rm off-shell}(\theta)\ =\
  \frac 32\sum_{m,m'=0,\pm}\rho_{mm}\,d^1_{mm'}(\theta)\,
  d^1_{mm'}(\theta)\,\,H_{m'm'}}\nonumber\\&&
  -\frac 32\,\Big(1-\frac{q^2}{m_W^2}\Big)\left(\rho_{t0}H_{t0}
  +\rho_{0t}H_{0t}\right)\cos\theta
  +\frac 32\,\Big(1-\frac{q^2}{m_W^2}\Big)^2\rho_{tt}H_{tt}.\qquad
\end{eqnarray}
The scalar analyzing functions $H_{t0}$, $H_{0t}$ and $H_{tt}$ are
proportional to the square of the quark masses and, therefore, the angular
distribution (\ref{offshell}) collapses back into the form (\ref{Wtheta}) in
the zero quark mass limit.

In the SM the Higgs couples to the gauge bosons via the metric tensor. The
resulting density matrix elements of the $W^\ast,Z^\ast$ bosons in the decay 
$H\to WW^\ast,ZZ^\ast$ have been calculated in~\cite{Groote:2013xt}.

\begin{figure}
\begin{center}
\epsfig{figure=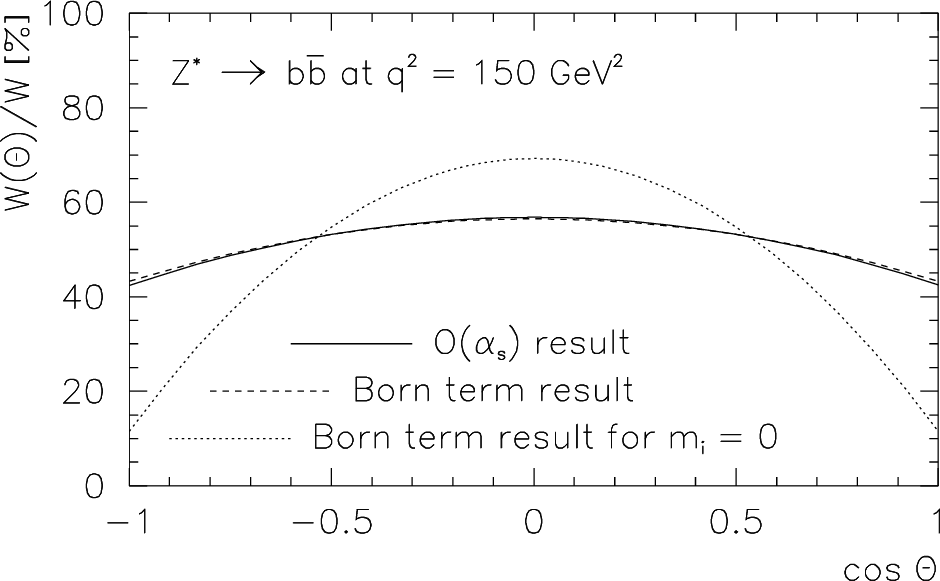, scale=0.8}
\end{center}
\caption{\label{nurkzbr}Polar angle distribution for $Z^{\ast}(\to b+\bar b)$
  at $q^2=150\,{\rm GeV^2}$. The three curves correspond to
  (i) Born term $(m_b=0)$ (dotted line)
  (ii) Born term $(m_b\neq 0)$ (dashed line) and
  (iii) $O(\alpha_s)$ with $(m_b\neq 0)$ (full line).}
\end{figure}

Numerical results for off-shell and quark mass effects in the decay 
$H\to W^-+W^{\ast +}(\to c\bar b)$ can be found in~\cite{Groote:2013xt}. Here
we concentrate on numerical results for the decay
$H\to Z+Z^\ast(\to b\bar b)$. At LO one finds that the scalar contribution not
present in the zero quark mass limit amounts to $8.6\,\%$ of the total decay
rate. In the vicinity of the threshold region the LO convexity parameter is
given by 
\begin{equation}\label{cfZZ}
c_f=-\frac 32\,\,\frac{q^2-4m_b^2}{q^2+2m_b^2}.
\end{equation}
The convexity parameter vanishes at threshold $q^2=4m_b^2$ as it must, whereas
one has $c_f=-3/2$ at $q^2=0$ in the zero mass limit. One anticipates big
differences of the angular decay distribution in the vicinity of the threshold
for the two cases. This shows up in the angular decay distribution
Fig.~\ref{nurkzbr} where we plot the angular decay distribution for
$q^2=150\,{\rm GeV}^2$ which is sufficiently far above the nonperturbative
regime of the $(b\bar b)$ channel.
The angular decay distribution is much flatter for the massive case.
Fig.~\ref{nurkzbr} also includes a plot of the NLO radiative corrections to
the massive case which can be seen to be quite small.  

\vspace{12pt}\noindent
{\bf Acknowledgements:}
This work was supported by the Estonian target financed project No.~0180056s09,
and by the Estonian Science Foundation under grant No.~8769. S.G.\
acknowledges support by the Forschungszentrum of the
Johannes-Gutenberg-Universit\"at Mainz ``Elementarkr\"afte und Mathematische
Grundlagen (EMG)''.

\end{document}